\documentclass[sigconf]{acmart}

\usepackage{booktabs}
\usepackage{array}
\usepackage{multirow}
\usepackage{amsmath}

\usepackage{amssymb}
\usepackage{algorithm}
\usepackage{algorithmic}
\usepackage{graphicx}
\usepackage{xcolor}

\setcopyright{none}
\acmConference[KDD '27]{Proceedings of the 33rd ACM SIGKDD Conference on
  Knowledge Discovery and Data Mining}{August 2027}{Location, Country}
\acmBooktitle{Proceedings of the 33rd ACM SIGKDD Conference on Knowledge
  Discovery and Data Mining (KDD '27)}
\renewcommand\footnotetextcopyrightpermission[1]{}
\newcommand{\symfootnote}[2]{\begingroup\renewcommand\thefootnote{#1}\footnotetext{#2}\endgroup}

\begin{document}

\title{Memory Layer: Train the In-Model Cache for Recommendation Models}

\author{Liangyuan Na\textsuperscript{*}, Gufan Yin\textsuperscript{*}, Yixin Bao\textsuperscript{*}, Xianjie Chen\textsuperscript{$\dagger$}, Justin Lin, Ziheng Huang, Xinyuan Zhang, Wen Zhang, Hao Lin, Xiaoheng Mao, Shuo Tang, Min Yu, Lei Chen, Chao Yang, Ziliang Zhao, Mengjiao Zhou, Zheng Qi, Dmitry Barablin, Chuo-Yun Yang, Kaustubh Vartak, Tingting Zhang, Arun Kumar Singh}
\affiliation{\institution{Meta}\country{\,}}

\renewcommand{\shortauthors}{Na, Yin, and Bao, et al.}

\begin{abstract}
Early ranking stages in recommendation systems precompute item embeddings and cache them in-model for scoring within strict latency constraints. Because this cache exists only at serving time, outside the training loop, training and serving use different item representations, a structural discrepancy that limits quality and adds operational fragility. We show that co-designing the training and serving paths removes this representation discrepancy at its source. We introduce the \emph{memory layer}, an in-model key-value embedding cache co-trained with the model: the item tower writes embeddings during training and the model reads them at serving, one source of truth for item representations by construction. Always-on embeddings cover items not yet cached, so every item receives a prediction, and the design consolidates three separate trainer-to-predictor update paths into a single self-contained pipeline. Deployed in production on Instagram Reels, the memory layer raises prediction coverage from 96\% to 100\%, improves embedding freshness from $O(5\text{ min})$ to $O(20\text{ s})$, and narrows the training-serving Normalized Entropy (NE) gap by up to 86\%, yielding over $2\times$ recall for the freshest content and a 5--6\% cold start engagement lift. Because embeddings are produced during training, the system needs no separate bulk-evaluation or publish-time recomputation, cutting training-and-publish computational cost by 30\% at neutral serving computational cost.
\end{abstract}




\maketitle
\pagestyle{fancy}
\fancyhf{}
\renewcommand{\headrulewidth}{0pt}
\fancyhead[L]{\itshape Memory Layer: Train the In-Model Cache for Recommendation Models}
\fancyhead[R]{\shortauthors}
\fancyfoot[C]{\thepage}
\symfootnote{*}{These authors contributed equally to this work.}
\symfootnote{$\dagger$}{Corresponding author (cxj@meta.com).}
\section{Introduction}
Modern recommendation systems at industrial scale employ multi-stage cascading architectures to balance relevance and computational efficiency. Early-stage ranking (ESR) occupies an important position in this cascade: it must score $O(1\text{K})$ candidate items per request with latency constraints measured in double-digit milliseconds, while providing sufficient ranking quality to surface relevant content for downstream stages. To meet these constraints, the dominant approach is a folded architecture where heavy item-side feature extraction is decoupled from request-time inference: an item tower precomputes dense embeddings offline, which are cached in-model for fast lookup during serving~\cite{silvertorch}. The two-tower paradigm~\cite{inttower,ebr}, where independent user and item towers produce embeddings scored via an interaction network, has become the standard for pre-ranking~\cite{inttower}. In this work, we \emph{co-train} the in-model cache with recommendation models, introducing the memory layer, which turns the read-only serving cache into a trained-and-streamed component that unifies the training and serving item representation into a single source of truth.
\subsection{The Conventional Approach}
\label{sec:conventional}
The in-model item embedding cache, introduced as part of the SilverTorch serving framework~\cite{silvertorch}, enables ESR models to precompute item embeddings and store them within the model's serving state. The model trains on training data from Scribe~\cite{scribe}, a real-time log-streaming service. The serving cache is populated from two separate sources. A GPU bulk evaluation service periodically processes the candidate pool (hundreds of millions of items) in Hive~\cite{hive}, an open-source data warehouse. It reads item features from the candidate pool table, backfilled with a multi-hour delay. A separate delta item streaming job consumes the candidate pool scribe to update newly created items at $O(5\text{ min})$ freshness. Items not present in the cache cannot be scored directly and require a system-specific fallback. Figure~\ref{fig:system} (a) illustrates this conventional architecture with two distinct training and serving paths. The system thus maintains three separate update paths from trainer to serving predictor: model parameter streaming/delta updates~\cite{quickupdate}, full snapshot bulk evaluation through the backfilled Hive table, and delta item streaming via a standalone job.
Each is an external dependency whose delay adds latency and operational burden and can surface as stale or missing embeddings at serving.
\subsection{Training-Serving Discrepancy}
\label{sec:discrepancy}
Despite its effectiveness, this system introduces structural training-serving discrepancies that limit model quality and create operational fragility. We identify three categories of inconsistency between what the model sees during training and what it encounters at serving:

\noindent\textbf{Item tower freshness.} During training, the model always uses the latest item tower with current parameters. At serving time, item embeddings are generated from an anchor checkpoint that may be hours old, and then cached. The serving model thus operates on stale item representations that lag behind the continuously-updated training state.

\noindent\textbf{Feature source discrepancy.} Training data contains the latest item features from the training data scribe at the time of logging. At serving time, item embeddings are computed from features in the candidate pool backfilled Hive table, a separate source with different logging configuration and hours of pipeline delay~\cite{feature-staleness}. This dual-source architecture introduces systematic feature discrepancies (e.g., missing or stale features for certain item types) that are difficult to detect offline, and causes the system to react slowly to trending content.

\noindent\textbf{Cache-miss behavior.} During training, the model never encounters a cache miss: the item tower is always executed. At serving time, an item not present in the cache is handled by one of three fallbacks: a fixed negative score that leaves it effectively unranked, deferral to a later-stage model with a random probability, or reliance on additional infrastructure to backfill the prediction. The model cannot learn to handle cold start items gracefully, and the effect of cache misses cannot be measured during offline development.

Together, these issues result in a measured 5--12\% gap between training Normalized Entropy (NE) and serving NE across engagement prediction tasks. \S\ref{sec:o2o} provides a systematic analysis of how the memory layer addresses each discrepancy.

\subsection{Memory Layer Approach: Training the In-Model Cache}
We observe that the cause is architectural: the item embedding cache exists only at serving time, outside the training loop. Training computes fresh embeddings via the item tower; serving looks up a separately-maintained cache. The two never share state. We propose the memory layer as a persistent, streamed item embedding cache that is co-trained with the model and directly serves predictions at inference time. By making the cache a component of both forward passes, writable in training, readable at serving, we remove the discrepancy at its source. This work makes four contributions:
\begin{enumerate}
\item \textbf{Model:} we introduce the memory layer, a co-trained sparse embedding table that provides one source of truth for item representations across training and serving, narrowing the training-serving NE gap by up to 86\%: from 12.11\% to 1.64\% on content selection (pselect), and from 5.24\% to 3.42\% on reshare. The in-model cache is then complemented with always-on embeddings to guarantee a prediction for every item.
\item \textbf{System:} we design and build the end-to-end infrastructure, a gradient-based Writeback mechanism, multi-table training for comprehensive cache population, Multi-Probe Zero-Collision Hashing (MPZCH)~\cite{mpzch} distributed serving with cache-miss handling, raw embedding streaming for near-real-time propagation, and a feature cache for auxiliary item signals, consolidating three separate update paths into a single self-contained pipeline.
\item \textbf{Benefit:} the unified design improves several dimensions at once: freshness from $O(5\text{ min})$ to $O(20\text{ s})$, 100\% cold start coverage (+5--6\% new-content breakout, +5--6\% cold start engagement lift), 30\% lower training-and-publish computational cost, improved serving reliability, and better training-serving fidelity.
\item \textbf{Validation:} we deploy on Instagram Reels, a surface of Instagram with a large number of daily active users, in production since late 2025 with consistent online benefits.
\end{enumerate}

\begin{figure*}[t]
  \centering
  \includegraphics[width=0.46\linewidth]{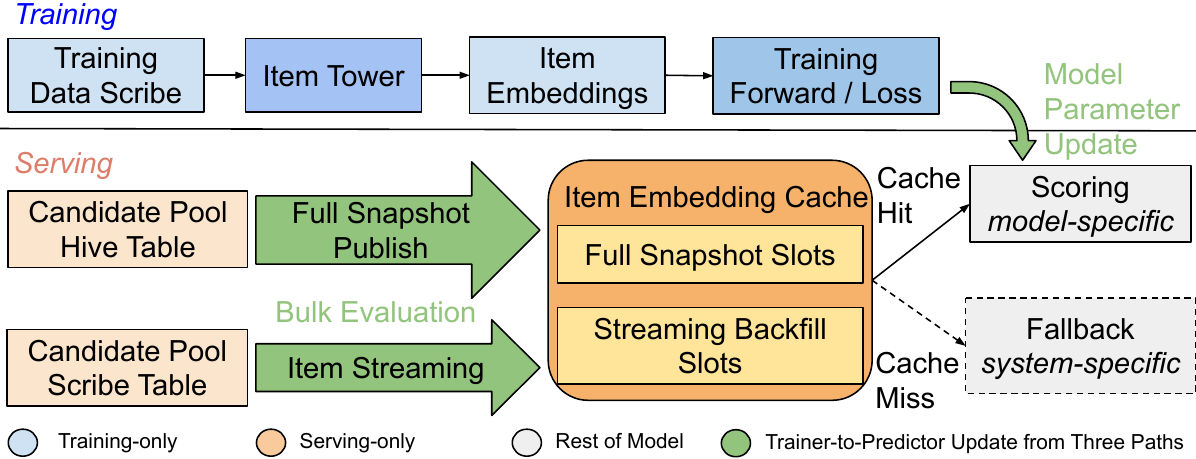}
  \hfill
  \includegraphics[width=0.46\linewidth]{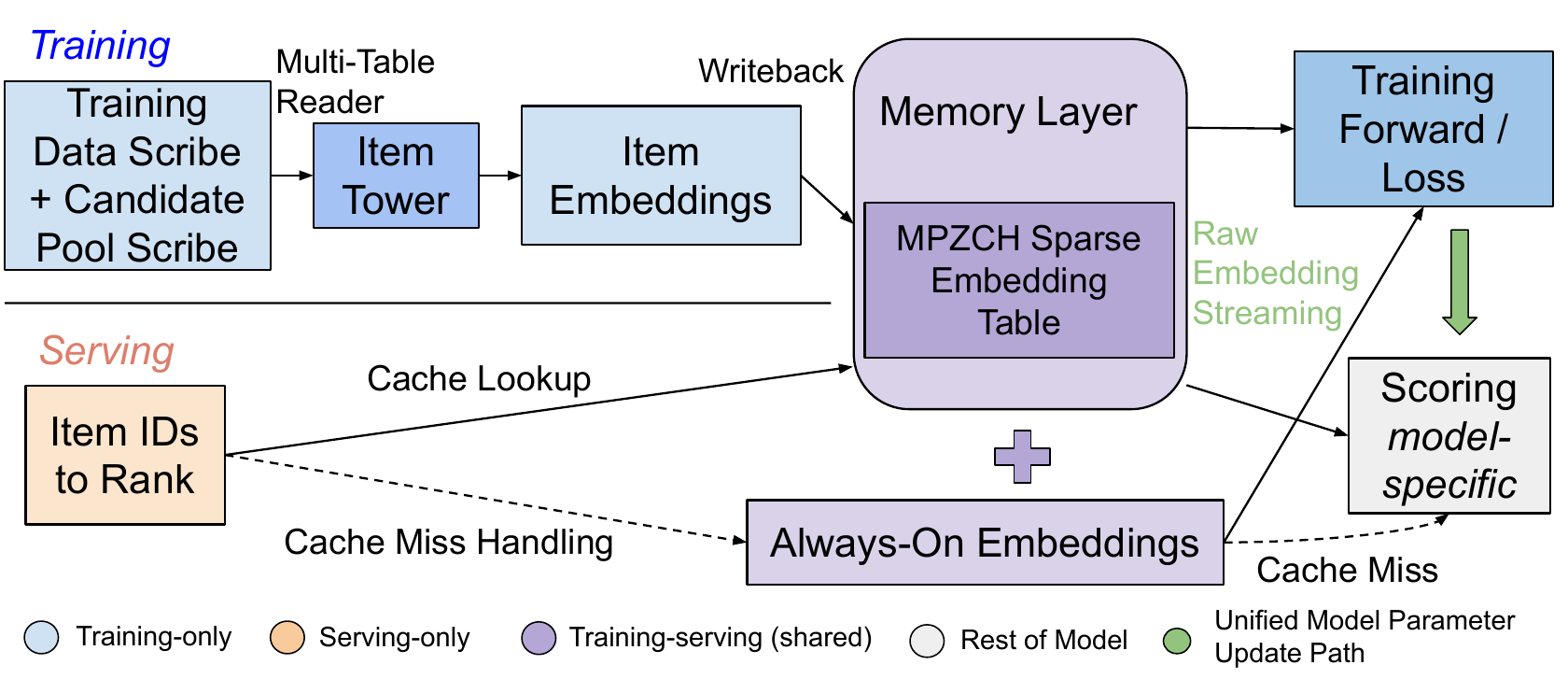}
  \caption{System comparison. (a) Conventional ESR system: training and serving are separate, with three trainer-to-predictor update paths and cache-missed items handled by a fallback. (b) Memory layer: training and serving share one co-trained embedding cache over a single update path, and always-on embeddings score every item, making the system self-contained.}
  \Description{Side-by-side system diagrams. (a) The conventional ESR system has three separate trainer-to-predictor paths: model parameter streaming, full-snapshot bulk evaluation through a backfilled Hive table, and a standalone item embedding streaming job. (b) The memory-layer system consolidates these into a single self-contained pipeline centered on a shared MPZCH embedding table that is written during training and streamed to the predictor.}
  \label{fig:system}
\end{figure*}

Figure~\ref{fig:system} (b) illustrates the resulting system architecture. The memory layer is a shared sparse embedding table using MPZCH~\cite{mpzch} that serves as the single source of truth for item representations in both training and serving. Multi-table training reads from both the training data scribe and candidate pool scribe directly, the item tower writes embeddings into the memory layer via Writeback, and raw embedding streaming propagates all updates to the predictor at 15-second intervals. The predictor performs MPZCH lookup with cache-miss handling via always-on embeddings, scoring 100\% of items. Compared to Figure~\ref{fig:system} (a), the three separate update paths are consolidated into a single self-contained pipeline centered on the shared memory layer.

\section{Related Work}
\label{sec:related}

\noindent\textbf{In-model item caches.} Industrial recommendation uses multi-stage
cascades~\cite{youtube,ranktower,dlrm} that decompose ranking into retrieval,
early-stage ranking, and final ranking. At datacenter scale, memory-bound
embedding-table operations account for a large amount of their inference computational cost~\cite{fb-rec-arch},
motivating in-model embedding caches for item representations. Two-tower
models~\cite{ebr,sbc,cold,inttower} factorize scoring into independent user
and item representations, enabling offline precomputation of item embeddings;
SilverTorch~\cite{silvertorch} internalizes this caching as an in-model
operator with GPU-accelerated lookup~\cite{cache-aware-rl}. Our memory layer
extends this in-model cache by making it \emph{writable during training},
turning a read-only serving artifact into a co-trained component; it is not
tied to the two-tower factorization and applies equally to unified rankers such
as SlimPer~\cite{slimper}. Its
embedding backbone (collision-free storage and distributed operators) builds
on MPZCH~\cite{mpzch} (\S\ref{sec:mpzch}) and TorchRec Table-Batched
Embedding (TBE)~\cite{torchrec}. Monolith~\cite{monolith} similarly pursues
collisionless tables, but we integrate the table into the training loop as a
cache.

\noindent\textbf{Challenges in serving cached embeddings.} Three challenges arise. First,
\emph{freshness}: online training~\cite{facebook-ads,ftrl} keeps models
current, but its delta publishing propagates parameter updates only at minute
scale, too slow for fresh item embeddings.
Methods that accelerate propagation, whether high-frequency delta publishing
(QuickUpdate~\cite{quickupdate}) or streaming parameter updates to inference
replicas (Ekko~\cite{ekko}), target model parameters rather than an item
cache.
Raw Embedding Streaming (RES) instead co-locates capture with the
training TBE prefetch pass and streams the co-trained item cache with no
dedicated infrastructure (\S\ref{sec:res}). Second, \emph{consistency}:
training-serving discrepancy is a known source of technical
debt in production systems~\cite{tech-debt,data-lifecycle}, yet feature stores~\cite{tfx} address
feature-level but not representation-level consistency. Third, \emph{cold start coverage}: new items must still be scored, yet prior cold start methods~\cite{coldstart-metrics,coldstart-meta} treat them as special cases. The memory layer addresses all three at once through a single shared table (\S\ref{sec:mpzch}).

\noindent\textbf{Memory-augmented networks.} Conceptually, the memory layer relates to memory-augmented
networks~\cite{ntm,dnc,memory-layers} in other domains, which use an external
key-value store of trainable parameters to target model capacity and reasoning
ability. We instead target training-serving consistency in recommendation
modeling, with bulk-write, real-time-streaming, and zero-collision
requirements.
\section{Memory Layer: Model Design}
\label{sec:model}

\subsection{Definition and Overview}
\label{sec:definition}
\noindent\textbf{Definition.} The memory layer $M$ is a key-value embedding
table integrated directly into the model architecture: the \emph{key} is the
item ID (media identifier), denoted $i$; the \emph{value} $M[i]$ is that item's cached embedding (a dense vector, the item tower output); and it is \emph{implemented} as a sparse
embedding table with collision-free hashing and least-recently-used (LRU) eviction.
During training the item tower writes embeddings into $M$; during serving the model reads them back for prediction, one source of truth across both phases, with the write-then-read mechanics detailed in \S\ref{sec:model-arch}.

\noindent\textbf{A co-trained cache.} The memory layer is not a static lookup but a \emph{co-trained} component: its rows are learned jointly with the rest of the model so the cached representation tracks the evolving item tower. There are two natural ways to learn it. The first is ordinary gradient descent, an $L_2$ objective $\lVert M[i]-\mathrm{sg}(e_i)\rVert_2^2$ (with $e_i$ the item tower output and $\mathrm{sg}(\cdot)$ the stop-gradient) that pulls the cached vector toward $e_i$ over successive steps. The second is \emph{Writeback} (\S\ref{sec:writeback}): a single-step exact assignment that overwrites $M[i]$ with the current item tower output every iteration (its target is the item tower output, not a task loss), reusing the model's existing optimizer path with no additional optimizer state. Both approaches produce comparable results (\S\ref{sec:ablation}), and we use Writeback for its exact assignment (\S\ref{sec:writeback}), which avoids tuning a loss weight and the transient staleness of gradual convergence.

\subsection{Model Architecture}
\label{sec:model-arch}
\begin{figure}[htbp]
  \centering
  \includegraphics[width=0.8\linewidth]{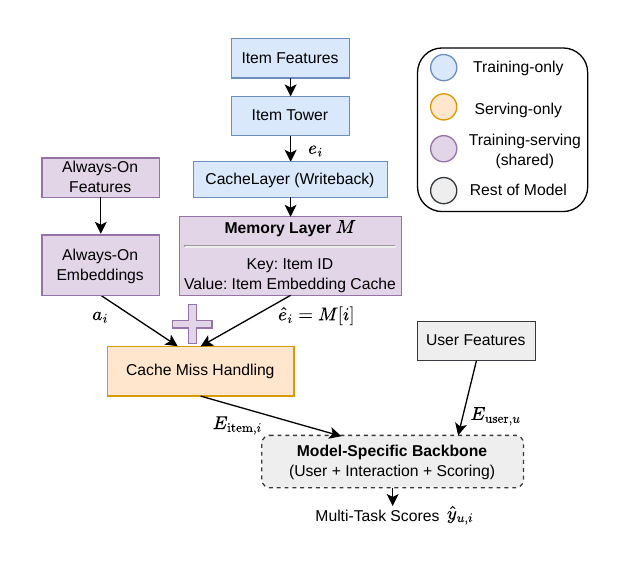}
  \caption{Unified model architecture with the memory layer.}
  \Description{Block diagram of the unified model. The item path runs item features through the item tower, writes the output into the shared memory layer (MPZCH) via CacheLayer, and reads it back. The item representation (memory layer output plus the always-on author-ID embedding) and the user representation feed the rest of the model architecture, which emits multi-task scores. The memory layer is shared across the training and serving paths.}
  \label{fig:model}
\end{figure}

Figure~\ref{fig:model} illustrates the unified model architecture centered on the memory layer. The model comprises an item path centered on the memory layer and the rest of the model architecture:

\noindent\textbf{Item path.} The item tower $f_\text{item}$ processes item features, i.e., features that are properties of the item itself rather than the request context, and produces dense item embeddings, denoted $e_i$. These embeddings are written into the memory layer $M$ via the CacheLayer Writeback mechanism. The model then reads back the cached value $\hat{e}_i = M[i]$, so the scoring network is trained on exactly what serving reads. The item side carries two independent gradient flows (\S\ref{sec:writeback}): Writeback sets $M[i]$ to the item tower output, and the ranking loss trains the model parameters, including the item tower. The two are decoupled, so the cache faithfully mirrors the item tower while the item tower is shaped by the task, and this write-then-read pattern keeps training aligned with inference.

\noindent\textbf{Model-specific backbone.} The remaining components are model-specific and operate independently of the memory layer. The user-side network produces a user representation $E_{\text{user},u}$ for user $u$ (e.g., from the tokenization results of the SlimPer framework~\cite{slimper}). A scoring network $g$ then combines $E_{\text{user},u}$ with $E_{\text{item},i}$ (from the item path: memory layer output plus always-on embeddings) to produce the multi-task scores $\hat{y}_{u,i} = g(E_{\text{user},u}, E_{\text{item},i})$; here $E_{\text{user},u}$ and $E_{\text{item},i}$ are ensemble embeddings, in contrast to the item embedding $e_i$ that the memory layer caches. We write $\theta$ for all model parameters. The memory layer is agnostic to this backbone, which can be any ranking architecture, such as SlimPer~\cite{slimper} or conventional two-tower models, and can evolve without affecting the item caching mechanism.

\noindent\textbf{Training vs.\ serving.} The two passes differ only on the item side: training runs the item tower and writes/reads $M$, while serving skips the item tower and reads $M$ directly, kept fresh via raw embedding streaming (RES, \S\ref{sec:res}), with cache misses handled by always-on embeddings $a_i$ (\S\ref{sec:always-on}).

\subsection{Always-On Embeddings}
\label{sec:always-on}
In conventional cached ESR systems, an item without a cached embedding cannot be scored directly and falls back to a fixed low score or deferral to a downstream stage, under-serving fresh content. We instead take a hybrid approach that splits the item-side features into two sets. The first set holds most of the item features: the memory layer caches the item-tower output computed from them, so at serving time the model reads this cached vector instead of extracting the features and rerunning the item tower, removing most of the per-item feature-extraction computational cost. The second set is a small number of features chosen to generalize to fresh items, such as the author ID, a topic ID, or other content-understanding features; these are still extracted at serving to produce \emph{always-on} embeddings. Because they describe an item through general attributes rather than its identity, they stay informative even for an item the model has never seen, whereas a per-item identity embedding such as the media ID is untrained for such an item.

During training the scoring network fuses the cached and always-on embeddings and learns how much to weight each per item: for fresher media the per-item cached signal is thin, so the model leans on the always-on embeddings, which stay meaningful regardless of item age. At serving time a cache hit uses both, while a cache miss zeroes the cached embedding and keeps only the always-on embeddings, so the representation degrades gracefully to the signal the model already relies on for fresh media. Every item therefore receives a reasonable prediction, replacing the previous behavior where cache-missed items lacked a proper score.

\subsection{Cache-based Evaluation}
\label{sec:cache-eval}
The cache-based evaluation functionality takes the memory layer as input and computes evaluation loss during training. This mirrors the serving logic, where cached embeddings are used for predictions, and helps close the training-serving development experience gap by allowing machine learning engineers to monitor metrics that closely reflect serving NE during offline development. In practice, during evaluation steps the model reads $\hat{e}_i = M[i]$ from the memory layer (identical to serving) rather than the fresh item tower output. If the memory layer contains stale or missing embeddings, the cache-based evaluation metric will reflect this, enabling engineers to diagnose training-serving issues during development rather than discovering them only after deployment. Combined with the memory layer's training-serving consistency, this mechanism contributes to narrowing the training-serving NE gap (\S\ref{sec:o2o}).

\section{Infrastructure Implementation}
\label{sec:infra}
Having defined the memory-layer model (\S\ref{sec:model}), we now detail the five infrastructure components that realize it,
with an additional feature-cache capability in Appendix~\ref{sec:feature-cache}. The components of MPZCH (\S\ref{sec:mpzch}), Writeback (\S\ref{sec:writeback}), and RES (\S\ref{sec:res}) have been released to the broader community through the open-source TorchRec~\cite{torchrec-repo} and FBGEMM~\cite{fbgemm-repo} libraries, while the model integration code remains internal.

\subsection{MPZCH Storage Backend}
\label{sec:mpzch}
The memory layer is implemented using Multi-Probe Zero-Collision Hashing (MPZCH)~\cite{mpzch} as its storage backend. MPZCH provides zero-collision ID-to-row mapping via multi-probe lookup, bounded capacity with LRU-based eviction, and GPU-optimized batched read/write operations via TorchRec TBE operators~\cite{torchrec}. The MPZCH module comprises three co-indexed components: (1) an identity tensor that maps raw item IDs to unique row indices via multi-probe lookup, residing on the GPU host for fast serving; (2) an embedding table storing item tower output vectors, sharded across GPU and CPU hosts via Distributed Inference (DI)~\cite{di-blog} for capacity scaling; and (3) a runtime metadata store for auxiliary item-side features (Appendix~\ref{sec:feature-cache}). All three share row indexing: the $i$-th row across all components corresponds to the same item, ensuring atomic updates. The table manages slot allocation through LRU eviction: when the bounded-capacity table (512-dimensional embeddings) is full, least-recently-accessed slots are reassigned to new items (see Appendix~\ref{sec:app-di} for DI sharding constraints and capacity analysis).

\subsection{CacheLayer: Writeback Algorithm}
\label{sec:writeback}
The CacheLayer is responsible for persisting item tower outputs into the memory layer during training. A naive approach, directly assigning values to embedding table rows, would bypass PyTorch's autograd system and require custom CUDA kernels for another distributed all-to-all communication. Instead, we design a Writeback mechanism that reuses the existing TBE backward infrastructure to achieve exact value assignment through gradient manipulation.

The intuition behind our Writeback mechanism is that the stochastic gradient descent (SGD) optimizer update rule, with learning rate $\eta{=}1$, can be repurposed as an exact assignment operator. Given the current cached value $\hat{e}_i$ and the target $e_i$ (the new item tower output), constructing the gradient as $g_i^\text{wb} = \hat{e}_i - e_i$ and applying EXACT\_SGD (the TorchRec sparse optimizer mode~\cite{torchrec} that performs vanilla SGD without momentum or weight decay) yields the in-place update
$\hat{e}_i \gets \hat{e}_i - 1 \cdot g_i^\text{wb} = e_i$. The update performs exact assignment without custom CUDA kernels, reusing the full TBE backward path (including distributed all-to-all for sharded tables). CacheLayer is a custom \texttt{torch.autograd.Function} that saves (current $-$ target) on the forward and returns it as the TBE gradient on the backward.

One property is \emph{gradient isolation}: the Writeback gradient updates $M$, while the training-loss gradient independently updates all model parameters $\theta$ (including the item tower), so $M$ faithfully reflects item tower outputs without interference from the training objective. Algorithm~\ref{alg:writeback} gives the full procedure, gradient construction, per-ID deduplication that keeps each ID's update exact, and the EXACT\_SGD step, with cross-rank and EmbeddingCollection (EC) index-dedup handling in Appendix~\ref{sec:app-dedup}.

\begin{algorithm}[t]
\caption{Memory Layer Writeback ($\mathbf{e}_\text{target}{=}e_i$, $\mathbf{e}_\text{cached}{=}\hat{e}_i$).}
\label{alg:writeback}
\begin{algorithmic}[1]
\REQUIRE Item IDs $\mathcal{I}$, item features $X_\mathcal{I}$, user representation $\mathbf{E}_\text{user}$, labels $Y$, Memory Layer $M$ (MPZCH, EXACT\_SGD, $\eta{=}1$), model params $\theta$
\ENSURE Updated $M$, training loss $\mathcal{L}$
\STATE \textbf{// Forward Pass}
\STATE $\mathbf{e}_\text{target} \gets f_\text{item}(X_\mathcal{I})$ \COMMENT{Item tower forward}
\STATE $\mathbf{e}_\text{cached} \gets M.\text{lookup}(\mathcal{I})$ \COMMENT{Current MPZCH values}
\STATE $\mathbf{g}_\text{wb} \gets \mathbf{e}_\text{cached} - \mathbf{e}_\text{target}$ \COMMENT{CacheLayer: save for backward}
\STATE $\mathcal{L} \gets L(g(\mathbf{E}_\text{user}, \mathbf{e}_\text{cached}), Y)$ \COMMENT{Loss on cached repr.}
\STATE \textbf{// Backward Pass}
\STATE $\mathcal{L}.\text{backward}()$ \COMMENT{Autograd for $\theta$}
\STATE \textbf{// CacheLayer returns $\mathbf{g}_\text{wb}$ as gradient to TBE}
\STATE \textbf{// TBE Optimizer Step (EXACT\_SGD, $\eta{=}1$)}
\FOR{each unique $i \in \mathcal{I}$}
    \STATE $M[i] \gets M[i] - 1 \cdot \mathbf{g}_\text{wb}[i] = M[i] - (M[i] -
  \mathbf{e}_\text{target}[i]) = \mathbf{e}_\text{target}[i]$ \COMMENT{Exact
  assignment}
  \ENDFOR
\STATE \textbf{// Result:} $M$ updated; $\nabla_{\theta}\mathcal{L}$ flows independently (decoupled)
\end{algorithmic}
\end{algorithm}

\subsection{Multi-Table Training}
\label{sec:multitable}
An important requirement is populating the memory layer with embeddings for the full candidate pool, not just items appearing in organic user interactions. In the conventional system (\S\ref{sec:conventional}), this is handled by a bulk evaluation service consuming the candidate pool Hive table, introducing a multi-hour pipeline delay. Multi-table training eliminates this dependency by reading both training data and candidate pool data directly into the online training loop via a multi-dataloader. The model receives two batches per iteration: a training batch (user interactions with labels) and a pool batch (candidate items without labels), with batch sizes matched to the throughput ratio between sources. The GPU preprocessing stage merges item features from both batches for efficient item tower computation. In the forward pass, the merged features go through the item tower and write embeddings to the memory layer. Embeddings are then split back: only training-batch items go through the scoring network for loss computation; pool items exist solely to populate the cache. Algorithm~\ref{alg:multitable} formalizes this procedure, with dummy-batch handling for asynchronous loaders in Appendix~\ref{sec:app-dummy}.

\begin{algorithm}[t]
\caption{Multi-Table Training with Memory Layer}
\label{alg:multitable}
\begin{algorithmic}[1]
\REQUIRE Training source $S_\text{train}$, Pool source $S_\text{pool}$, Memory Layer $M$, Item Tower $f_\text{item}$, model params $\theta$
\ENSURE Updated $M$, training loss $\mathcal{L}$
\STATE $B_\text{train}, B_\text{pool} \gets S_\text{train}.\text{next}(), S_\text{pool}.\text{next}()$ \COMMENT{async load}
\STATE \textbf{// Forward: merge, single item tower pass, Writeback}
\STATE $\mathbf{X}_\text{merged} \gets \text{concat}(\text{items}(B_\text{train}), \text{items}(B_\text{pool}))$ \COMMENT{GPU preproc}
\STATE $\mathbf{e}_\text{merged} \gets f_\text{item}(\mathbf{X}_\text{merged})$ \COMMENT{one forward for all items}
\STATE CacheLayer.forward($\mathbf{e}_\text{merged}$, $M.\text{lookup}(\mathcal{I}_\text{merged})$) \COMMENT{Writeback}
\STATE \textbf{// Loss on training batch only}
\STATE $\mathcal{L} \gets L(g(\mathbf{E}_\text{user}, \mathbf{E}_\text{item}), Y_\text{train})$ \COMMENT{Loss on training batch}
\STATE \textbf{// Backward: updates $\theta$ and, via Algorithm~\ref{alg:writeback}, $M$}
\STATE $\mathcal{L}.\text{backward}()$
\end{algorithmic}
\end{algorithm}

\subsection{Model Publish and Serving}
\label{sec:serving}
In conventional ESR systems, item embeddings are served from a dedicated embedding cache populated during bulk evaluation on a separate GPU cluster. The memory layer replaces this with a direct MPZCH lookup: the same table written during training is packaged into a serving artifact during model publish~\cite{silvertorch}. Since embeddings are already computed during training, publish reduces to checkpoint conversion, cutting publish time and computational cost (quantified in \S\ref{sec:reliability}). Full snapshot publishes occur periodically for fault tolerance; real-time freshness is maintained via raw embedding streaming (\S\ref{sec:res}).

Figure~\ref{fig:serving} (Appendix~\ref{sec:app-di}) illustrates the serving-time architecture. In the current deployment, the always-on set is the author ID, complemented on a cache hit by a per-item media-ID embedding. Unhashed item IDs are sent to the MPZCH identity tensor on the GPU host, which maps each ID to a row index and produces a \texttt{found\_mask} indicating cache hits. Remapped indices are distributed to remote DI shards for embedding retrieval. Simultaneously, the always-on author-ID table provides an identity embedding for every item regardless of cache status, while the media-ID lookup is gated by the cache-hit mask. The \texttt{found\_mask} gates the MPZCH and media-ID outputs: cache-hit items receive their full embedding; cache-missed items are zeroed, leaving the author-ID signal. The gated outputs are concatenated to form $E_{\text{item},i}$, passed to the scoring network.

\noindent\textbf{Cache-miss handling.} By default MPZCH returns an embedding from the first probed position for unknown IDs, indistinguishable from a valid lookup and liable to inflate cold-item scores. We instead make the identity tensor return a length-0 entry for cache-missed IDs, producing all-zero rows; the resulting \texttt{found\_mask} gates the MPZCH and media-ID outputs to zero, leaving only the author-ID signal (\S\ref{sec:always-on}). DI sharding constraints and observability are detailed in Appendix~\ref{sec:app-di}.

\subsection{Raw Embedding Streaming (RES)}
\label{sec:res}
Raw Embedding Streaming (RES) propagates memory layer updates from the online trainer to the serving predictor in near-real-time, reducing latency from $O(5\text{ min})$ to $O(20\text{ s})$ with 100\% row coverage (Figure~\ref{fig:res}, Appendix~\ref{sec:app-res}).

The design insight is co-locating streaming with the TBE prefetch I/O pass: RES captures updated embeddings while rows are already resident in GPU high-bandwidth memory (HBM) for the next training iteration, eliminating a separate copy step. The pipeline operates as a 3-stage asynchronous process. Stage 1 (every iteration): stores lookup IDs and pulls updated rows from HBM to CPU asynchronously. Stage 2 (continuous, non-blocking): deduplicates by ID, applies quantization, and merges across shards in a separate process. Stage 3 (every 15 seconds): serializes and pushes batched updates to the predictor via the messaging infrastructure. RES streams three data types atomically: MPZCH embedding rows, identity tensor updates, and runtime metadata / feature cache (Appendix~\ref{sec:feature-cache}). All three stages are non-blocking; CPU-side processing does not block the GPU training loop. This unified architecture eliminates the standalone item streaming job (previously a dedicated GPU cluster), removing a separate serving dependency while achieving 21-second P99 end-to-end latency (detailed pipeline design in Appendix~\ref{sec:app-res}).

\section{Reliability and Efficiency Analysis}
\label{sec:reliability}
\subsection{System Simplification and Reliability}
\label{sec:simplification}
The memory layer consolidates multiple disjoint components into a unified architecture, reducing
operational complexity in three ways. \textbf{Simplified serving infrastructure:}
training and serving logic are aligned by construction, eliminating external dependencies on
bulk evaluation entirely; both full-snapshot and item-streaming bulk evaluations are removed,
leaving the online trainer as the sole source of item representations. \textbf{Streamlined
services:} two previously separate streaming services (sparse parameter and item embedding
streaming) merge into one infrastructure with two channels (RES for embeddings, sparse for other
parameters), and two embedding-maintenance mechanisms (the SilverTorch cache module and
separate sparse tables) unify into a single MPZCH-managed TBE table. \textbf{Minimized external
dependencies:} reliance on the candidate-pool Hive table, which carries a multi-hour
pipeline delay, is reduced, and the standalone item streaming job is eliminated. This
simplicity is itself a reliability property: fewer moving parts and dependencies mean fewer
points whose delay or unavailability can surface as stale or missing embeddings, hardening the
serving path. Consistent with this,
the memory-layer system has run in production since late 2025, improving
both serving reliability and developer experience over the conventional architecture.

\subsection{Resource Efficiency}
The self-contained design removes the dedicated item-streaming infrastructure (its work absorbed
into the online trainer via RES) and reduces publish to a lightweight checkpoint conversion,
cutting full-snapshot publish time and delivering $3\times$ higher
recurrent-publish throughput, while freed hosts are reallocated to online training.
Overall, the
memory layer achieves 30\% lower training-and-publish computational cost per model, aggregating across all models that use it.
On the serving side, inference-throughput optimizations (Appendix~\ref{sec:app-throughput}) improve end-to-end serving throughput by $2.42\times$, matching the production baseline's serving computational cost while delivering better model quality. The net operating point is thus lower training and publish computational cost at neutral serving computational cost, plus the reliability (\S\ref{sec:simplification}) and cold start (\S\ref{sec:coldstart}) gains.

\section{Experimental Results}
\label{sec:results}

\subsection{Experimental Setup}
We deploy and evaluate the memory layer on Instagram Reels at the early-stage ranking layer, serving a large number of daily active users. The results reported here are from a two-tower instantiation of the memory layer~\cite{ebr,inttower}. The baseline model uses the SilverTorch in-model embedding cache~\cite{silvertorch} (\S\ref{sec:conventional}), whose capacity is bounded by the predictor's GPU serving memory, and bypasses uncached items with a fixed negative score.
The deployed system adds the memory-layer components of \S\ref{sec:model}--\S\ref{sec:infra} with feature cache (Appendix~\ref{sec:feature-cache}) as a further extension.

Populated during training as a TorchRec TBE table with flexible sharding for distributed inference~\cite{torchrec} (\S\ref{sec:mpzch}, Appendix~\ref{sec:app-di}), the memory layer can scale its table size and embedding dimension as needed.
The memory layer's serving-efficiency improvements (\S\ref{sec:reliability}) let us increase the shared table capacity by 50\% over the baseline at neutral serving computational cost, so the larger capacity is part of the coverage improvement rather than an unequal comparison.

We report NE~\cite{facebook-ads}, cache hit rate, embedding time-to-serve (TTS), engagement rates (reshare, time spent, video views), and cold start metrics (breakout rate, recency, and creation-to-first-impression time, i.e., CTI).

\subsection{Cache Hit and Embedding Freshness}
\label{sec:cachehit}
\begin{figure}[t]
  \centering
  \includegraphics[width=\linewidth]{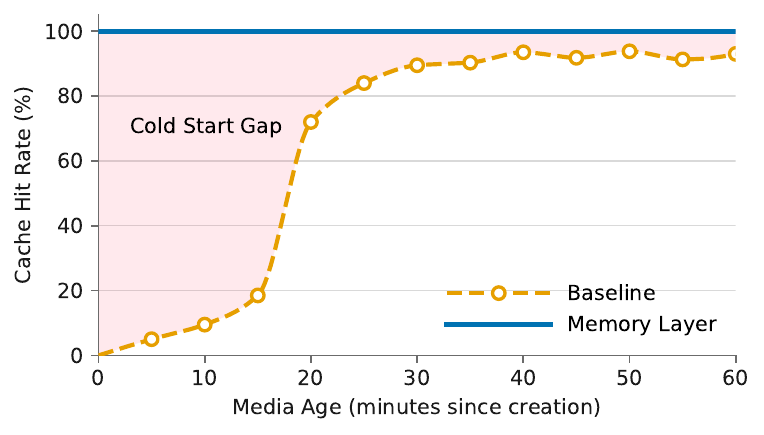}
  \caption{Cache hit rate versus media age.}
  \Description{Line chart of cache hit rate versus media age. The baseline curve falls below 20 percent for content under 15 minutes old and recovers as age increases, while the memory-layer curve stays flat at 100 percent across all media ages.}
  \label{fig:cachehit}
\end{figure}

Overall prediction coverage increases from 96\% (baseline) to 100\% (memory layer). The MPZCH memory layer achieves 99.5\% cache hit rate by itself, with the remaining 0.5\% of previously-missed items handled by always-on embeddings. In the baseline system, cache hit rate was only 10--20\% for media created within the last 15 minutes; after the memory layer, cache availability is fully agnostic to media age (Figure~\ref{fig:cachehit}). Reaching 100\% coverage is what makes the ESR stage self-contained: conventional systems rely on a fallback for uncached items, whereas the memory layer scores every item, so item scoring is fully owned within the stage. The same property makes the approach directly applicable to final-stage ranking, where every candidate must receive a score.

Embedding freshness improves from $O(5\text{ min})$ to $O(20\text{ s})$. We measure time-to-serve (TTS) as the end-to-end latency from training forward pass to predictor patching. The P99 TTS is approximately 21 seconds, primarily the 15-second streaming interval (component breakdown in Appendix~\ref{sec:app-res}). The memory layer also eliminates full-snapshot cache overrides: in the baseline system, each hourly full-snapshot deployment overwrites recently-streamed item embeddings with staler representations computed from a delayed Hive table, causing a periodic recency drop visible in online metrics. With a single source of item embeddings from training to predictor, this production artifact is eliminated; the model maintains stable recency throughout the full-snapshot cycle.

\subsection{Cold Start Performance}
\label{sec:coldstart}
Since deployment, the memory layer improves cold start through five complementary mechanisms: (1) faster streaming ($O(5\text{ min})\!\to\!O(20\text{ s})$); (2) continuous online training that removes the full-snapshot override problem; (3) expanded cache capacity (by 50\%) with multi-source ingestion; (4) always-on embeddings that predict for cache-missed items; and (5) streaming from the latest item tower rather than an hours-old checkpoint. Together these drive both recall improvements (primarily driven by fresh-content cache-hit increase in \S\ref{sec:cachehit}) and model prediction quality improvements for cold start content.

In an A/B backtest after launch, cold start recall improves video views by 6--7\% for items with 1-hour recency and by over $2\times$ for media created within 5 minutes. Cold start quality also improves across the cold start pool, with statistically significant engagement-rate improvements of 5--6\% in reshare and time spent. These improvements translate to attributed topline impact of 5--6\% in breakout rate (the fraction of new items reaching 10{,}000 views within 3 days) and in the 1-hour recency metric, along with a 40--45\% reduction in P25 CTI, meaning the fastest-served quartile of new content reaches its first impression sooner (Table~\ref{tab:coldstart}). An A/B pretest ablation confirms these improvements are attributable to the memory layer rather than concurrent model changes.

\begin{table}[t]
\caption{Summary of cold start performance improvements. All A/B improvements are statistically significant.}
\label{tab:coldstart}
\small
\setlength{\tabcolsep}{4pt}
\begin{tabular}{@{}llc@{}}
\toprule
Category & Metric & Improvement \\
\midrule
\multirow{2}{*}{Recall (A/B)} & Video views (media age $<$1h) & $+6$--$7\%$ \\
 & Video views (media age $<$5 min) & $>2\times$ \\
\midrule
\multirow{2}{*}{Quality (A/B)} & Reshare rate & $+5$--$6\%$ \\
 & Time spent & $+5$--$6\%$ \\
\midrule
\multirow{3}{*}{Topline} & Breakout (10K/3d) & $+5$--$6\%$ \\
 & 1-hour recency & $+5$--$6\%$ \\
 & P25 CTI & $-40$--$45\%$ \\
\bottomrule
\end{tabular}
\end{table}

\subsection{Training-Serving Consistency}
\label{sec:o2o}
The memory layer resolves each discrepancy of \S\ref{sec:discrepancy}: item tower freshness via online training and RES (\S\ref{sec:res}), the dual feature source via multi-table training on the same scribe (\S\ref{sec:multitable}), and cache misses via always-on embeddings (\S\ref{sec:always-on}).

By design, training NE and serving NE share the same item representations: a single source of truth (the MPZCH embedding table) is used in both training and inference. In offline iterations, training NE is computed on the same cached embeddings used for serving. This makes the training-serving consistency of early-stage ranking equivalent to that of late-stage models which do not use external caches. We track the gap between online training NE and online serving NE in Table~\ref{tab:o2o-gap} for two representative tasks: pselect (content selection scoring, predicting which items are selected by later ranking stages) and reshare (reshare-propensity prediction). The training-serving gap is measured as $(\text{NE}_\text{serve} - \text{NE}_\text{train}) / \text{NE}_\text{train} \times 100\%$, with both metrics measured at the same timestamp window to control for diurnal patterns.

Sharing one item representation shrinks these gaps: the pselect gap drops by 86\% (12.11\% to 1.64\%) and the reshare gap by 35\% (5.24\% to 3.42\%). The residual gap is small but nonzero because sharing the table removes the representation discrepancy, not every training-serving difference. We attribute it to three sources: (i) int8 row-wise quantization applied at serving but not in training; (ii) write-behind reads, where serving uses embeddings written a few steps earlier (\S\ref{sec:discussion}); and (iii) cache-miss fusion, which the model learns only implicitly during training: fresher items, whose cached embedding is less informative, approximate a cache miss, but this implicit pattern differs in distribution from the explicit all-zero miss encountered at serving. Improving this fusion is future work (\S\ref{sec:discussion}). We therefore report an 86\% reduction, not full elimination.

\begin{table}[t]
\caption{Training-serving NE gap for two representative tasks.}
\label{tab:o2o-gap}
\small
\setlength{\tabcolsep}{5pt}
\begin{tabular}{@{}lccc@{}}
\toprule
Task & Baseline Gap & Memory Layer Gap & Gap Reduction \\
\midrule
pselect & 12.11\% & 1.64\% & 86\% \\
reshare & 5.24\% & 3.42\% & 35\% \\
\bottomrule
\end{tabular}
\end{table}

\subsection{Ablation Study}
\label{sec:ablation}
The components are integrated, so strict single-component ablations of the deployed system would require infrastructure changes. We instead report the qualitative directional contribution of each component, added
on top of the previous configuration (Table~\ref{tab:ablation}). We note these measurements come from
an early prototype, and each component may differ slightly from the latest production implementation. Thus they are only used to validate our design decisions during development and are not directly comparable to the results reported
elsewhere in this section for exact contribution.
The co-trained MPZCH cache with Writeback establishes training-serving NE
consistency and engagement wins over the bulk-evaluation baseline; always-on embeddings, MPZCH
lookup serving, and raw embedding streaming each add further engagement and cold start improvements.
Multi-table training with cache scale-up makes the system self-contained with broader coverage;
and feature cache achieves parity results with additional infra capability. A component-level
comparison of the two cache-learning objectives (\S\ref{sec:definition}) finds Writeback and the
$L_2$ variant comparable, with Writeback slightly better on pselect NE, so we choose Writeback
for its exactness.

\begin{table}[htbp]
\caption{Component-wise ablation from early prototype.}
\label{tab:ablation}
\small
\setlength{\tabcolsep}{4pt}
\renewcommand{\arraystretch}{1.1}
\begin{tabular}{@{}>{\raggedright\arraybackslash}p{3.6cm}>{\raggedright\arraybackslash}p{4.5cm}@
{}}
\toprule
Configuration & A/B metrics improved \\
\midrule
MPZCH and Writeback & Reshare, time spent \\
Always-on embeddings & Cold start views and time spent \\
MPZCH lookup serving & Reshare, time spent, views \\
Raw Embedding Streaming & Reshare \\
\bottomrule
\end{tabular}
\end{table}

\subsection{Evaluation Considerations}
\label{sec:eval-considerations}
\noindent\textbf{Media distribution.} A factor worth discussing separately is how the data helps shape what the ablation measures. The training and candidate-pool sources control which media populate the cache. Reaching full coverage therefore changes which items are scored: the memory layer now ranks the ${\sim}4\%$ of items the baseline bypassed, so
treatment and control no longer see identical media pools. This is a
serving-time exposure change, not a training-gradient effect: pool items are unlabeled and do not enter the loss. Because cold start media has lower immediate engagement, short-horizon A/B under-reads the benefit; we therefore rely on controlled serving evaluation on identical media sets and longer-running experiments. We also vary the data and feature source and the cache size: the framework already yields improvements without these changes, while multi-table training offers the flexibility to further adjust the media distribution.

\noindent\textbf{Evaluation scope and reproducibility.} Our evaluation is confined to a single production surface (Instagram Reels early-stage ranking) and uses internal data with no public counterpart, so exact reproduction is not possible. Cross-system comparison is also hard: the memory layer integrates into each surface's model and serving stack, and infrastructures differ in data sources, cache sizes, and serving constraints. The design nonetheless uses standard building blocks (collision-free hashing, an $\eta{=}1$ SGD write, and raw embedding streaming) detailed in \S\ref{sec:model}--\S\ref{sec:res} to support reimplementation.

\section{Discussion}
\label{sec:discussion}
\subsection{Limitations}
\noindent\textbf{Write-behind staleness.} Writeback is inherently write-behind: if item $i$ is last processed at training step $t$, the cache stores $M[i]=f_\text{item}(x_i;\theta_t)$, but at serving time $t+k$ the model weights have advanced to $\theta_{t+k}$ while serving still reads the stale $M[i]$. The resulting gap grows with $k$, the number of training steps since item $i$ was last refreshed. In practice this is a large improvement, not a regression: the conventional system refreshes items at $O(5\text{ min})$ to $O(1$--$2\text{ h})$, whereas the memory layer streams a single path at $O(20\text{ s})$, down to $O(\text{seconds})$ for frequently-trained items, and eliminates the slow bulk evaluation path. Multi-table training (\S\ref{sec:multitable}) bounds $k$ by cycling all candidate-pool items through training, so even rarely-served items refresh periodically. The residual staleness is inherent to any write-behind cache, a lower bound absent real-time item tower inference at serving.

\noindent\textbf{Cache-miss handling.} During training the item tower computes every embedding, so cache misses never enter the loss and the model is never explicitly taught how to fuse the cached and always-on embeddings when the cached one is absent; this makes cache-miss fusion a source of the residual training-serving NE gap (\S\ref{sec:o2o}). Learning the fusion explicitly is ongoing work along three directions: (i) simulating the serving-time cache-miss pattern during training, e.g., by having the MPZCH backend emit a training-time miss mask that mirrors serving, so the model learns to fuse under the distribution it meets at inference; (ii) enriching the always-on set with more item-agnostic features so a cache-missed item retains more signal; and (iii) an architectural treatment that makes the fusion robust to a missing component, drawing on the missing-modality literature in multimodal learning~\cite{smil,mmrobust}.

\subsection{Lessons Learned}
Three lessons generalize beyond this deployment. \emph{Consolidation results in reliability}: collapsing three trainer-to-predictor paths into one removed several classes of external dependency (\S\ref{sec:reliability}), a larger reliability improvement than any single-component tuning. \emph{Coverage changes measurement}: reaching 100\% coverage shifts the served media distribution toward cold start content, which typically draws lower immediate engagement than established content, so short-horizon A/B under-reads the benefit and controlled serving evaluation is required (\S\ref{sec:ablation}-\S\ref{sec:eval-considerations}). \emph{Exactness beats tuning}: we prefer exact mechanisms (the $\eta{=}1$ Writeback and kernel-level runtime-metadata design, Appendix~\ref{sec:feature-cache}) over the gradient-loss and second-table variants we first tried, which needed loss-weight tuning or corrupted int64 feature values.

\subsection{Generalization and Future Work}
\label{sec:discussion-future}
The memory layer's principles extend beyond this deployment. Because it is a general-purpose, model-co-located key-value store, the same mechanism supports: (i) \emph{cross-stage generalization}: with always-on embeddings it spans early-stage (cache-everything) to late-stage (cache-nothing) ranking, and a natural extension is a single model serving both stages; (ii) \emph{generalized caching}: the runtime-metadata design (Appendix~\ref{sec:feature-cache}) can hold arbitrary features or cached history embeddings, reducing dependence on feature stores; and (iii) \emph{LLM foundation-model integration}: a pre-trained language or
vision-language model writes item embeddings into the memory layer during
training. Because the cache is co-trained with the model
(\S\ref{sec:definition}), whether by Writeback or the $L_2$ variant, ranking
gradients can propagate back into the foundation model to fine-tune or
co-train it end-to-end. Serving reads only the cached embeddings, incurring
\emph{zero} foundation-model inference computational cost. This pattern makes otherwise-prohibitive foundation models deployable
within a strict-latency ESR stage. It also connects to memory-augmented
networks~\cite{memory-layers}, whose addressing ideas (learned eviction, hierarchical memory, product-key lookup) could inform future designs.

\section{Conclusion}
\label{sec:conclusion}
We presented the memory layer, a model-serving co-design that turns the item embedding cache from a read-only serving artifact into a \emph{co-trained} component. Making the same sparse table writable in training, readable at serving, and streamed between the two in seconds establishes one source of truth for item representations, removing the training-serving representation discrepancy by construction, while always-on embeddings guarantee a prediction for every item.

Deployed on Instagram Reels since late 2025, it reaches 100\% prediction coverage and
improves embedding freshness from $O(5\text{ min})$ to $O(20\text{ s})$, while narrowing the
training-serving NE gap by up to 86\%. These improvements translate to
topline impact: over $2\times$ recall for the freshest content and a 5--6\% cold start engagement lift. The self-contained design also delivers a 30\% reduction in training-and-publish computational cost at neutral serving computational cost and improves serving reliability from a simpler system.

More broadly, co-designing the training and serving paths yields quality, reliability, and efficiency improvements that are hard to obtain by optimizing either path alone. The principles extend to cross-stage ranking, generalized caching, and LLM foundation-model integration, positioning the in-model cache as a foundation for next-generation recommendation systems.

\begin{acks}
\noindent\textbf{Engineering collaborators.} We are grateful to collaborators across Meta whose discussion and support were invaluable to this work: Yiming Liao, Xiaochen Hou, Tuan Trieu, Su Min Kim, Rui Jian, Xialu Li, Faran Ahmad, Sarunya Pumma, Supadchaya Puangpontip, Albert Chen, Enzhou Liu, Ji Liu.

\noindent\textbf{Leadership.} We are also grateful for the leadership support of Misael Manjarres, Lan Gao, William Pei, Zhengyu Su, Guangdeng Liao, Haotian Wu, Deepak Agarwal, Peng Xia, Yiyi Pan, Bi Xue, Shilin Ding, whose guidance and investment made this work possible.
\end{acks}

\bibliographystyle{ACM-Reference-Format}
\bibliography{references}

@inproceedings{smil,
  author    = {Ma, Mengmeng and Ren, Jian and Zhao, Long and Tulyakov, Sergey and Wu, Cathy and Peng, Xi},
  title     = {{SMIL}: Multimodal Learning with Severely Missing Modality},
  booktitle = {Proceedings of the 35th AAAI Conference on Artificial Intelligence (AAAI)},
  publisher = {AAAI Press},
  year      = {2021},
}

@inproceedings{mmrobust,
  author    = {Ma, Mengmeng and Ren, Jian and Zhao, Long and Testuggine, Davide and Peng, Xi},
  title     = {Are Multimodal Transformers Robust to Missing Modality?},
  booktitle = {Proceedings of the IEEE/CVF Conference on Computer Vision and Pattern Recognition (CVPR)},
  publisher = {IEEE},
  year      = {2022},
}

@article{hive,
  author    = {Thusoo, Ashish and Sarma, Joydeep Sen and Jain, Namit and Shao, Zheng and Chakka, Prasad and Anthony, Suresh and Liu, Hao and Wyckoff, Pete and Murthy, Raghotham},
  title     = {Hive: A Warehousing Solution over a Map-Reduce Framework},
  journal   = {Proceedings of the VLDB Endowment},
  volume    = {2},
  number    = {2},
  pages     = {1626--1629},
  year      = {2009},
}

@inproceedings{silvertorch,
  author={Xue, Bi and Wu, Hong and Chen, Lei and Yang, Chao and Ma, Yiming and Ding, Fei and Wang, Zhen and Wang, Liang and Mao, Xiaoheng and Huang, Ke and others},
  title     = {{SilverTorch}: A Unified Model-based System to Democratize Large-Scale Recommendation on {GPUs}},
  booktitle = {Proceedings of the 49th International ACM SIGIR Conference on Research and Development in Information Retrieval (SIGIR)},
  publisher = {ACM},
  year      = {2026},
}

@inproceedings{memory-layers,
  author = {Berges, Vincent-Pierre and O{\u{g}}uz, Barlas and Haziza, Daniel and Yih, Wen-tau and Zettlemoyer, Luke and Ghosh, Gargi},
  title = {Memory Layers at Scale},
  booktitle = {Proceedings of the 42nd International Conference on Machine Learning},
  publisher = {PMLR},
  year = {2025},
}

@article{mpzch,
  author={Zhao, Ziliang and Xue, Bi and Lin, Emma and Lu, Tianqi and Zhou, Mengjiao and Vartak, Kaustubh and Ali-Zade, Shakhzod and Li, Tao and Kuang, Bin and Jian, Rui and others},
  title   = {Multi-Probe Zero Collision Hash ({MPZCH}): Mitigating Embedding Collisions and Enhancing Model Freshness in Large-Scale Recommenders},
  journal = {arXiv preprint arXiv:2602.17050},
  year    = {2026},
}

@inproceedings{inttower,
  author={Li, Xiangyang and Chen, Bo and Guo, HuiFeng and Li, Jingjie and Zhu, Chenxu and Long, Xiang and Li, Sujian and Wang, Yichao and Guo, Wei and Mao, Longxia and others},
  title     = {{IntTower}: The Next Generation of Two-Tower Model for Pre-Ranking System},
  booktitle = {Proceedings of the 31st ACM International Conference on Information and Knowledge Management (CIKM)},
  publisher = {ACM},
  year      = {2022},
}

@inproceedings{ebr,
  author    = {Huang, Jui-Ting and Sharma, Ashish and Sun, Shuying and Xia, Li and Zhang, David and Pronin, Philip and Padmanabhan, Janani and Ottaviano, Giuseppe and Yang, Linjun},
  title     = {Embedding-based Retrieval in {Facebook} Search},
  booktitle = {Proceedings of the 26th ACM SIGKDD International Conference on Knowledge Discovery \& Data Mining (KDD)},
  publisher = {ACM},
  year      = {2020},
}

@inproceedings{feature-staleness,
  author  = {Wang, Zhikai and Shen, Yanyan and Zhang, Zibin and Lin, Kangyi},
  title   = {Feature Staleness Aware Incremental Learning for {CTR} Prediction},
  booktitle = {Proceedings of the Thirty-Second International Joint Conference on Artificial Intelligence (IJCAI)},
  publisher = {International Joint Conferences on Artificial Intelligence Organization},
  year    = {2023},
}

@inproceedings{youtube,
  author    = {Covington, Paul and Adams, Jay and Sargin, Emre},
  title     = {Deep Neural Networks for {YouTube} Recommendations},
  booktitle = {Proceedings of the 10th ACM Conference on Recommender Systems (RecSys)},
  publisher = {ACM},
  year      = {2016},
}

@inproceedings{ranktower,
  author    = {Yan, YaChen and Li, Liubo},
  title     = {{RankTower}: A Synergistic Framework for Enhancing Two-Tower Pre-Ranking Model},
  booktitle = {Proceedings of the AdKDD Workshop at KDD},
  publisher = {ACM},
  year      = {2024},
}

@article{dlrm,
  author={Naumov, Maxim and Mudigere, Dheevatsa and Shi, Hao-Jun Michael and Huang, Jianyu and Sundaraman, Narayanan and Park, Jongsoo and Wang, Xiaodong and Gupta, Udit and Wu, Carole-Jean and Azzolini, Alisson G and others},
  title   = {Deep Learning Recommendation Model for Personalization and Recommendation Systems},
  journal = {arXiv preprint arXiv:1906.00091},
  year    = {2019},
}

@inproceedings{sbc,
  author    = {Yi, Xinyang and Yang, Ji and Hong, Lichan and Cheng, Derek Zhiyuan and Heldt, Lukasz and Kumthekar, Aditee and Zhao, Zhe and Wei, Li and Chi, Ed},
  title     = {Sampling-Bias-Corrected Neural Modeling for Large Corpus Item Recommendations},
  booktitle = {Proceedings of the 13th ACM Conference on Recommender Systems (RecSys)},
  publisher = {ACM},
  year      = {2019},
}

@inproceedings{cold,
  author    = {Wang, Zhe and Zhao, Liqin and Jiang, Biye and Zhou, Guorui and Zhu, Xiaoqiang and Gai, Kun},
  title     = {{COLD}: Towards the Next Generation of Pre-Ranking System},
  booktitle = {Proceedings of the DLP-KDD Workshop at KDD},
  publisher = {ACM},
  year      = {2020},
}

@inproceedings{cache-aware-rl,
  author  = {Chen, Xiaoshuang and Zhang, Gengrui and Wang, Yao and Wu, Yulin and Su, Shuo and Zhan, Kaiqiao and Wang, Ben},
  title   = {Cache-Aware Reinforcement Learning in Large-Scale Recommender Systems},
  booktitle={Companion Proceedings of the ACM Web Conference 2024},
  publisher = {ACM},
  year={2024}
}

@inproceedings{recd,
  author={Zhao, Mark and Choudhary, Dhruv and Tyagi, Devashish and Somani, Ajay and Kaplan, Max and Lin, Sung-Han and Pumma, Sarunya and Park, Jongsoo and Basant, Aarti and Agarwal, Niket and others},
  title     = {{RecD}: Deduplication for End-to-End Deep Learning Recommendation Model Training Infrastructure},
  booktitle = {Proceedings of Machine Learning and Systems (MLSys)},
  publisher = {mlsys.org},
  year      = {2023},
}

@inproceedings{torchrec,
  author    = {Ivchenko, Dmytro and van der Staay, Dennis and Taylor, Colin and Liu, Xing and Feng, Will and Kindi, Rahul and Sudarshan, Anirudh and Sefati, Shahin},
  title     = {{TorchRec}: A {PyTorch} Domain Library for Recommendation Systems},
  booktitle = {Proceedings of the 16th ACM Conference on Recommender Systems (RecSys)},
  publisher = {ACM},
  year      = {2022},
}

@inproceedings{monolith,
  author = {Liu, Zhuoran and Zou, Leqi and Zou, Xuan and Wang, Caihua and Zhang, Biao and Tang, Da and Zhu, Bolin and Zhu, Yijie and Wu, Peng and Wang, Ke and others},
  title   = {Monolith: Real Time Recommendation System With Collisionless Embedding Table},
  booktitle={Proceedings of 5th Workshop on Online Recommender Systems and User Modeling, jointly with the 16th ACM Conference on Recommender Systems},
  publisher = {ACM},
  year={2022}
}

@article{ntm,
  author  = {Graves, Alex and Wayne, Greg and Danihelka, Ivo},
  title   = {Neural Turing Machines},
  journal = {arXiv preprint arXiv:1410.5401},
  year    = {2014},
}

@article{dnc,
  author={Graves, Alex and Wayne, Greg and Reynolds, Malcolm and Harley, Tim and Danihelka, Ivo and Grabska-Barwi{\'n}ska, Agnieszka and Colmenarejo, Sergio G{\'o}mez and Grefenstette, Edward and Ramalho, Tiago and Agapiou, John and others},
  title   = {Hybrid Computing Using a Neural Network with Dynamic External Memory},
  journal = {Nature},
  publisher = {Nature Publishing Group},
  volume  = {538},
  number  = {7626},
  year    = {2016},
}

@inproceedings{facebook-ads,
  author={He, Xinran and Pan, Junfeng and Jin, Ou and Xu, Tianbing and Liu, Bo and Xu, Tao and Shi, Yanxin and Atallah, Antoine and Herbrich, Ralf and Bowers, Stuart and others},
  title     = {Practical Lessons from Predicting Clicks on Ads at {Facebook}},
  booktitle = {Proceedings of the 8th International Workshop on Data Mining for Online Advertising (AdKDD)},
  publisher = {ACM},
  year      = {2014},
}

@inproceedings{ftrl,
  author={McMahan, H Brendan and Holt, Gary and Sculley, David and Young, Michael and Ebner, Dietmar and Grady, Julian and Nie, Lan and Phillips, Todd and Davydov, Eugene and Golovin, Daniel and others},
  title     = {Ad Click Prediction: A View from the Trenches},
  booktitle = {Proceedings of the 19th ACM SIGKDD International Conference on Knowledge Discovery and Data Mining (KDD)},
  publisher = {ACM},
  year      = {2013},
}

@inproceedings{tech-debt,
  author    = {Sculley, D. and Holt, Gary and Golovin, Daniel and Davydov, Eugene and Phillips, Todd and Ebner, Dietmar and Chaudhary, Vinay and Young, Michael and Crespo, Jean-Fran{\c{c}}ois and Dennison, Dan},
  title     = {Hidden Technical Debt in Machine Learning Systems},
  booktitle = {Advances in Neural Information Processing Systems (NeurIPS)},
  publisher = {Curran Associates, Inc.},
  year      = {2015},
}

@article{data-lifecycle,
  author  = {Polyzotis, Neoklis and Roy, Sudip and Whang, Steven Euijong and Zinkevich, Martin},
  title   = {Data Lifecycle Challenges in Production Machine Learning: A Survey},
  journal = {SIGMOD Record},
  publisher = {ACM},
  year    = {2018},
}

@inproceedings{tfx,
  author={Baylor, Denis and Breck, Eric and Cheng, Heng-Tze and Fiedel, Noah and Foo, Chuan Yu and Haque, Zakaria and Haykal, Salem and Ispir, Mustafa and Jain, Vihan and Koc, Levent and others},
  title     = {{TFX}: A {TensorFlow}-Based Production-Scale Machine Learning Platform},
  booktitle = {Proceedings of the 23rd ACM SIGKDD International Conference on Knowledge Discovery and Data Mining (KDD)},
  publisher = {ACM},
  year      = {2017},
}

@inproceedings{coldstart-metrics,
  author    = {Schein, Andrew I. and Popescul, Alexandrin and Ungar, Lyle H. and Pennock, David M.},
  title     = {Methods and Metrics for Cold-Start Recommendations},
  booktitle = {Proceedings of the 25th Annual International ACM SIGIR Conference on Research and Development in Information Retrieval (SIGIR)},
  publisher = {ACM},
  year      = {2002},
}

@inproceedings{coldstart-meta,
  author    = {Wang, Li and Jin, Binbin and Huang, Zhenya and Zhao, Hongke and Lian, Defu and Liu, Qi and Chen, Enhong},
  title     = {Preference-Adaptive Meta-Learning for Cold-Start Recommendation},
  booktitle = {Proceedings of the 30th International Joint Conference on Artificial Intelligence (IJCAI)},
  publisher = {International Joint Conferences on Artificial Intelligence Organization},
  year      = {2021},
}

@article{gesr,
  author  = {Hong, Juhee and Liu, Meng and Wang, Shengzhi and Mao, Xiaoheng and Cheng, Huihui and Gao, Leon and Leung, Christopher and Zhou, Jin and Sekar, Chandra Mouli and Zhu, Zhao and Liu, Ruochen and Trieu, Tuan and Sun, Dawei and Kanjani, Jeet and Li, Rui and Qian, Jing and Cao, Xuan and Fan, Minjie and Gao, Mingze},
  title   = {Generative Early Stage Ranking},
  journal = {arXiv preprint arXiv:2511.21095},
  year    = {2025},
}

@misc{di-blog,
  author       = {Fang, Lu and Deng, Shiyan and Jia, Hongyi and Li, Huamin and Mitra, Ilina and Qin, Sheng and Zhang, Zhengkai and Zhao, Zhuoran and Zheng, Zinnia},
  title        = {Building Highly Efficient Inference System for Recommenders Using {PyTorch}},
  howpublished = {\url{https://pytorch.org/blog/building-highly-efficient-inference-system-for-recommenders-using-pytorch/}},
  year         = {2026},
  note         = {PyTorch Blog},
}

@inproceedings{quickupdate,
  author    = {Matam, Kiran Kumar and Ramezani, Hani and Wang, Fan and Chen, Zeliang and Dong, Yue and Ding, Maomao and Zhao, Zhiwei and Zhang, Zhengyu and Wen, Ellie and Eisenman, Assaf},
  title     = {{QuickUpdate}: A Real-Time Personalization System for Large-Scale Recommendation Models},
  booktitle = {21st USENIX Symposium on Networked Systems Design and Implementation (NSDI '24)},
  publisher = {USENIX Association},
  year      = {2024},
}

@article{slimper,
  author={Wang, Siqi and Chen, Xianjie and Deng, Shaofeng and Chen, Albert and Shah, Romil and Huang, Jiawei and Wang, Zhaoqin and Zhang, Zhang and Liu, Yiqun and Jiang, Meilei and others},
  title   = {{SlimPer}: Make Personalization Model Slim and Smart},
  journal = {arXiv preprint arXiv:2607.12281},
  year    = {2026},
}

@article{scribe,
  author  = {Karpathiotakis, Manos and Rizopoulos, Vlassios and Kahveci, Basri and Carotti, Tiziano and Gelum, Artem and Nada, Hazem and Dolgov, Yuri},
  title   = {{Scribe}: How {Meta} Transports Terabytes per Second in Real Time},
  journal = {Proceedings of the VLDB Endowment},
  publisher = {VLDB Endowment},
  year    = {2025},
}

@inproceedings{ekko,
  author    = {Sima, Chijun and Fu, Yao and Sit, Man-Kit and Guo, Liyi and Gong, Xuri and Lin, Feng and Wu, Junyu and Li, Yongsheng and Rong, Haidong and Aublin, Pierre-Louis and Mai, Luo},
  title     = {{Ekko}: A Large-Scale Deep Learning Recommender System with Low-Latency Model Update},
  booktitle = {16th USENIX Symposium on Operating Systems Design and Implementation (OSDI)},
  publisher = {USENIX Association},
  year      = {2022},
}

@inproceedings{fb-rec-arch,
  author={Gupta, Udit and Wu, Carole-Jean and Wang, Xiaodong and Naumov, Maxim and Reagen, Brandon and Brooks, David and Cottel, Bradford and Hazelwood, Kim and Hempstead, Mark and Jia, Bill and others},
  title     = {The Architectural Implications of {Facebook}'s {DNN}-based Personalized Recommendation},
  booktitle = {IEEE International Symposium on High-Performance Computer Architecture (HPCA)},
  publisher = {IEEE},
  year      = {2020},
}

@misc{torchrec-repo,
title        = {{TorchRec}},
author       = {{Meta Platforms, Inc.}},
howpublished = {\url{https://github.com/meta-pytorch/torchrec/}},
year         = {2026},
note         = {Software repository, accessed 2026},
}

@misc{fbgemm-repo,
title        = {{FBGEMM}},
author       = {{Meta Platforms, Inc.}},
howpublished = {\url{https://github.com/pytorch/FBGEMM}},
year         = {2026},
note         = {Software repository, accessed 2026},
}

\appendix
\section{Additional Infrastructure Details}
This appendix elaborates the infrastructure components of \S\ref{sec:infra}, introduces the feature cache extension, notes inference throughput optimization, and discusses deployment considerations.

\subsection{Writeback and EC Index Deduplication Interaction}
\label{sec:app-dedup}
The memory layer uses EmbeddingCollection with index deduplication enabled, which collapses duplicate indices within the local batch before all-to-all communication to reduce both communication volume and TBE lookup count~\cite{torchrec}. After the embedding lookup and output all-to-all, the full output is reconstructed by expanding deduplicated embeddings via \texttt{index\_select}. In backward, this expansion uses \texttt{scatter\_add} to reconstruct gradients, accumulating contributions for duplicate indices rather than keeping them separate. Without correction, the Writeback gradient magnitude for a frequently-appearing ID would be inflated proportionally to its batch count, violating the exact-assignment property. To handle this, CacheLayer computes the within-batch frequency of each ID during forward and normalizes the Writeback gradient by dividing by the corresponding frequency before the TBE optimizer step. This guarantees that the total Writeback for a given ID equals (cached $-$ target) regardless of how many times it appears. Per-ID deduplication (\S\ref{sec:writeback}) remains necessary for cross-rank duplicates, since index dedup operates only within the local batch on each rank.

\subsection{Multi-Table Dummy-Batch Handling}
\label{sec:app-dummy}
Since both dataloaders must keep pace with real-time data, we use timeout-based asynchronous loading. If either source lacks ready data, the system substitutes a pre-saved batch template (captured from the first valid batch after startup) as a dummy, maintaining correct tensor dimensions without contributing to model updates. Per-sample indicators control downstream behavior: dummy training batches have zeroed loss weights; dummy pool batches skip Writeback. This guarantees the training loop never stalls. Dummy templates are persisted across restarts for deterministic behavior, and the system logs dummy-batch frequency as a health metric to detect sustained data availability issues.

\subsection{DI Sharding and Cache-Miss Detection}
\label{sec:app-di}
\begin{figure}[htbp]
  \centering
  \includegraphics[width=\linewidth]{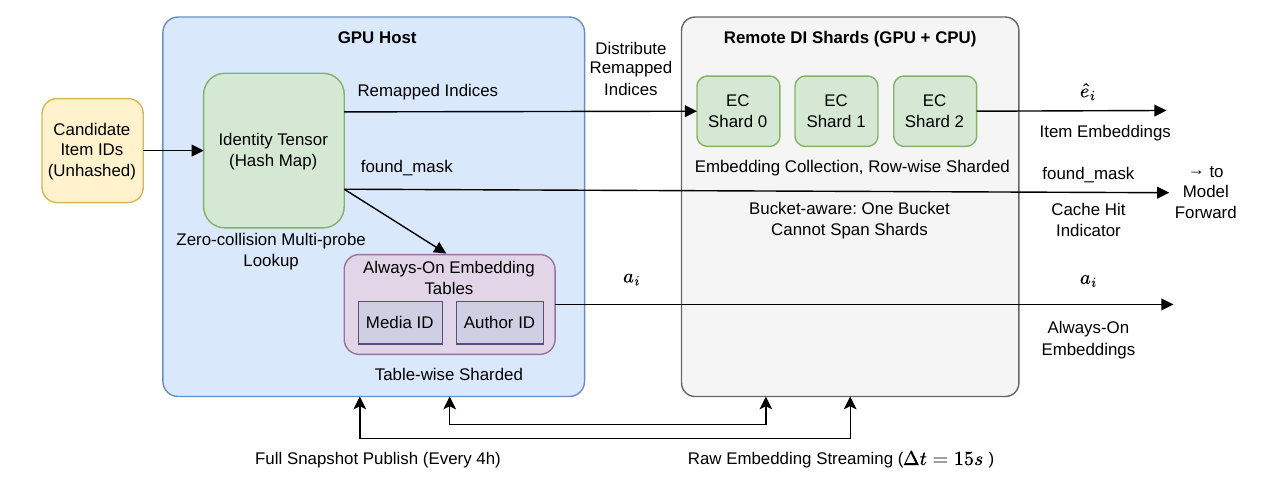}
  \caption{MPZCH Distributed Inference serving diagram.}
  \Description{Serving-time architecture diagram. Unhashed item IDs enter the MPZCH identity tensor on the GPU host, which emits row indices and a found-mask; the indices fan out to row-wise-sharded embedding-collection DI hosts. Always-on tables are table-wise sharded on the GPU host. The found-mask gates cache-missed rows to zero before they are concatenated into the item representation.}
  \label{fig:serving}
\end{figure}

\noindent\textbf{DI sharding.} To increase the memory-layer capacity by 50\% while maintaining inference latency, the MPZCH table is sharded across GPU and CPU hosts via Distributed Inference~\cite{di-blog}. The identity tensor resides on the GPU host for fast ID-to-row lookup, while the Embedding Collection is row-wise sharded across DI hosts for capacity. An important constraint is bucket-aware sharding: a single MPZCH bucket cannot be split across different DI shards, because probing occurs at the bucket level during ID lookup. The sharding strategy respects bucket boundaries to ensure correct lookup semantics. Always-on embedding tables use table-wise sharding (each table on one host) for better queries-per-second (QPS) given their small size. All other unused item-side embedding tables are trimmed to prevent latency increases from unnecessary sparse architecture output.

\noindent\textbf{Cache-miss detection.} By default, the MPZCH module returns an embedding from the first probed position for IDs not present in the identity tensor, indistinguishable from a valid lookup. This means cache-missed items would receive embeddings belonging to unrelated (often popular) items, inflating their scores. We implement explicit cache-miss detection: for cache-missed IDs, the identity tensor returns a length-0 entry in the remapped ID list, which indicates all-zero rows after the embedding forward and masking. The \texttt{found\_mask} tensor (derived from entry lengths) gates MPZCH and media ID outputs via element-wise masking. Cache hit rate is logged as a semi-binary value per item and aggregated in the serving metrics system, enabling per-pool analysis by joining with downstream metadata.

\subsection{Raw Embedding Streaming: Detailed Design}
\label{sec:app-res}
\begin{figure}[htbp]
  \centering
  \includegraphics[width=\linewidth]{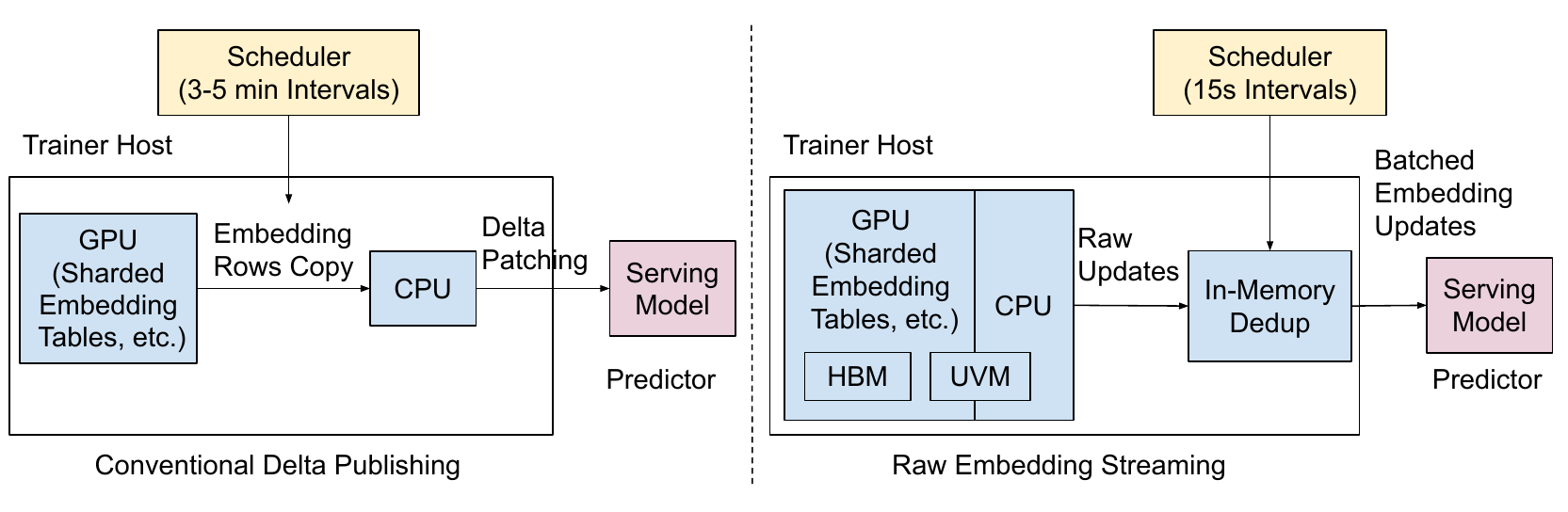}
  \caption{Comparison of conventional delta publishing (left) and the raw embedding streaming pipeline (right).}
  \Description{Two-part figure. Left: conventional delta publishing pipeline, which uses sync copy. Right: raw embedding streaming which fetches updated rows from GPU HBM and UVM.}
  \label{fig:res}
\end{figure}

\noindent\textbf{Motivation.} Low-latency model-update systems propagate parameter updates from the trainer to serving, by high-frequency delta publishing~\cite{quickupdate} or by streaming to replicas~\cite{ekko}. Conventional delta publishing~\cite{quickupdate} copies embedding deltas to a GPU memstore every iteration, then at 3--5 minute intervals copies weight rows from GPU to CPU for publish. This approach has drawbacks for unified virtual memory (UVM) caching or key-value (SSD) tables, as it triggers large batch copies of rows that may already be offloaded to DRAM or Flash, pausing the trainer. RES differs in two ways: (1) it captures updated embeddings during the existing TBE prefetch I/O pass, while rows are already resident in GPU HBM, eliminating a separate copy step; and (2) it runs as part of the training process itself with no dedicated streaming infrastructure.

\noindent\textbf{Iteration timing.} In Stage 1, RES stores the lookup IDs from iteration $t$ and, in iteration $t+1$, pulls the updated rows corresponding to those IDs from the TBE while they are still in HBM from the prefetch pass. The rows are asynchronously copied to CPU and sent to an in-memory store running in a separate process. This one-iteration delay ensures rows are captured after the Writeback update has been applied, guaranteeing consistency.

\noindent\textbf{Unified streaming.} Because item-embedding updates now flow through RES inside the online-training job, they combine with standard sparse streaming into a single trainer-to-predictor update path for all parameters (\S\ref{sec:res}).

\noindent\textbf{Latency breakdown.} The P99 end-to-end TTS (${\sim}21$ s) is primarily the streaming interval, with smaller contributions from embedding capture, deduplication, streaming preparation, message delivery, and predictor patching.

\subsection{Feature Cache}
\label{sec:feature-cache}
\begin{figure}[htbp]
  \centering
  \includegraphics[width=\linewidth]{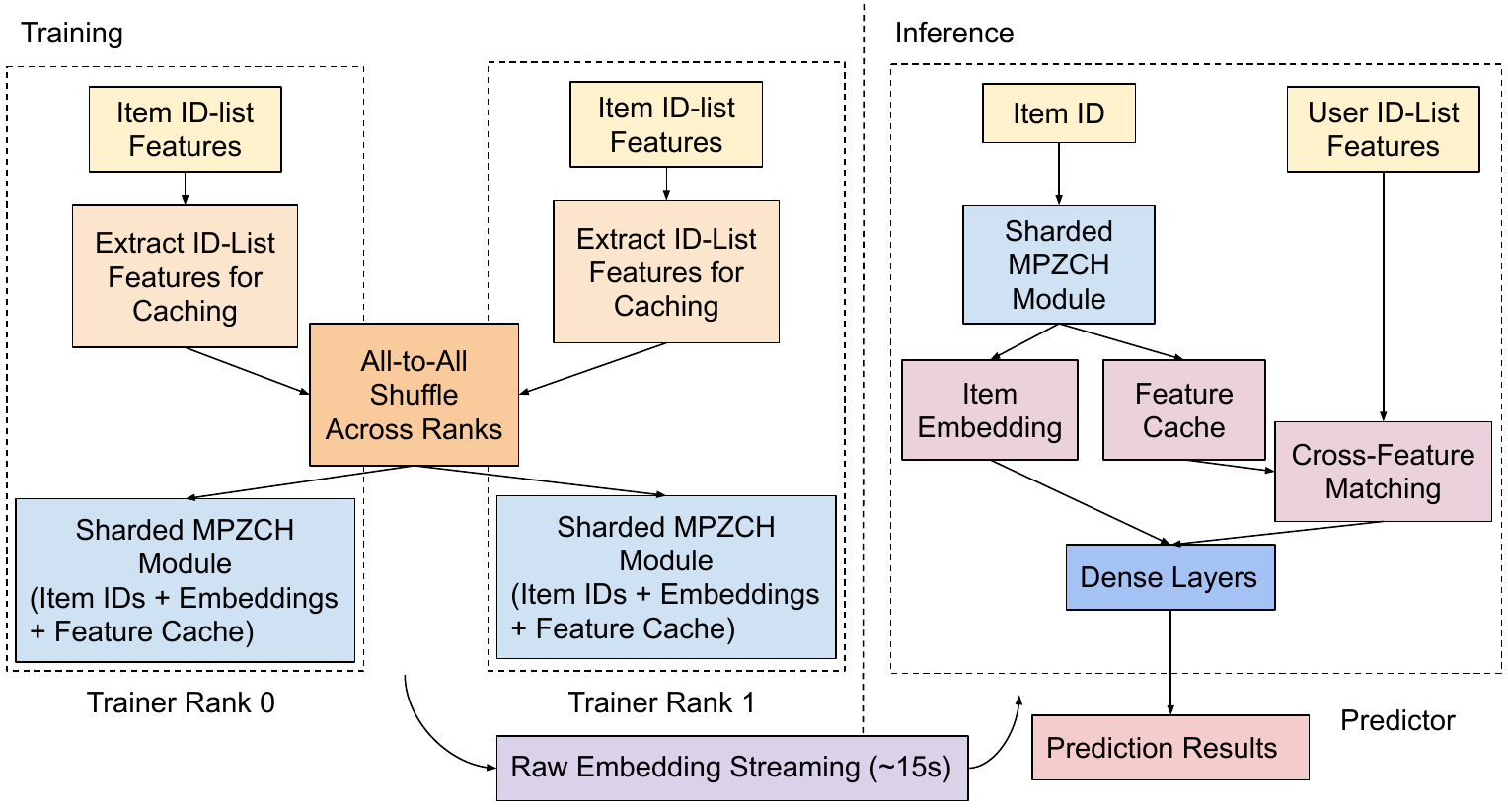}
  \caption{Feature cache end-to-end flow.}
  \Description{End-to-end feature-cache flow. Item-side raw features are written during training to co-located runtime metadata that shares the MPZCH identity-tensor indexing, and are streamed atomically together with the embeddings to the predictor, where they are looked up by item ID at serving time.}
  \label{fig:featurecache}
\end{figure}

Beyond item tower embeddings, ranking models often require item-side raw features at serving time. For example, explicit cross-feature matching~\cite{gesr} counts overlapping IDs between user engagement history and item-side features such as the content author's identifier. Without a feature cache, these must be fetched live from external services at serving time. We extend the memory layer to store these features alongside item embeddings. We generate and persist the feature cache during training, together with item IDs and item embeddings. In training, after reading ID-list features~\cite{recd}, we shuffle IDs and features all-to-all across ranks and persist them into the corresponding sharded MPZCH module. During inference, cached features are looked up by item ID from the same MPZCH table and used to compute downstream signals. In online training, all updated IDs and feature cache are streamed together to the predictor at ${\sim}15\text{s}$ cadence. Figure~\ref{fig:featurecache} illustrates the end-to-end flow.

The design must simultaneously satisfy three hard constraints: zero tolerance for precision loss (int64 feature values where any corruption causes a measurable NE regression), no additional computational overhead, and training-inference consistency. We extend the MPZCH module with a 2D parameter of shape $[N \times \text{num\_features}]$, where $N$ is the MPZCH table's row capacity, to store custom runtime metadata alongside the embedding table. The metadata shares the same indexing as the identity tensor: the feature cache for the $i$-th item is stored in the $i$-th row. After an item ID is successfully inserted into the MPZCH kernel, its features are written at the same slot; evicted slots are automatically zeroed. This atomic co-location eliminates synchronization issues between embeddings and features. During training, cross-feature values are encoded losslessly (int64 $\to$ float via bitwise reinterpretation) and written to MPZCH metadata slots via the existing all-to-all communication path with no additional computational overhead (encoding details below). At inference time, cached features are retrieved via index lookup on the runtime metadata, co-located with the identity tensor on the GPU host. Cache-missed features are masked to zero. Since item IDs and features are saved as a pair, they share the same cache hit rate with no additional serving latency. The feature cache is streamed atomically alongside embeddings via RES (\S\ref{sec:res}), ensuring no window where a predictor has a new item ID but stale features.

\noindent\textbf{Training encoding detail.} During training, cross-feature values are extracted from item-side ID-list features, encoded from int64 to float via bitwise reinterpretation (\texttt{torch.view}, lossless), and attached as 2D weights to the embedding input KeyedJaggedTensor (KJT)~\cite{torchrec, recd}. These are distributed via the existing all-to-all communication path. After MPZCH deduplication and ID insertion, features are written only at successfully inserted slots and decoded back to int64 before saving.

\noindent\textbf{Streaming atomicity.} One requirement is atomicity in RES: whenever an item ID delta is streamed, its corresponding feature cache must be included in the same update. We extended three streaming pipeline components (the post-lookup tracker, raw IDs tracker, and embedding kernel) to co-capture and co-stream runtime metadata, ensuring no window exists where a predictor has a new item ID but stale features.

\noindent\textbf{Alternatives considered.} We chose the runtime metadata design over two alternatives: (1) separate MPZCH modules (rejected due to unsynchronizable race conditions between independent hash maps and additional CUDA synchronization computational overhead), and (2) a second EmbeddingCollection with shared identity tensor using CacheLayer Writeback (rejected due to NaN canonicalization corrupting int64 values, MPZCH collision semantics, EC index dedup precision loss, and UVM cache staleness). The chosen design operates at the MPZCH kernel level, below autograd and TBE abstraction, where ID assignment and metadata writes are guaranteed atomic.

\subsection{Inference Throughput Optimization}
\label{sec:app-throughput}
The memory layer's structural property, each lookup returns exactly one embedding per item (pooling factor $=1$), enables the adoption of non-pooled EmbeddingCollection, unlocking index deduplication~\cite{torchrec}, uint8 sparse output quantization, and proportional core-table sharding~\cite{di-blog}, delivering ${\sim}40\%$ throughput improvement for the sparse component. Always-on embedding tables are placed on the GPU host via table-wise sharding, eliminating cross-host communication. Combined with parallelism tuning, these yield the end-to-end $2.42\times$ serving throughput reported in \S\ref{sec:reliability}.

\subsection{Operating Requirements and Deployment Trade-offs}
\label{sec:app-operating}
The memory layer depends on continuous online training to refresh item embeddings; systems that rely solely on periodic batch retraining cannot reach the $O(20\text{ s})$ freshness that closes the training-serving gap, so the approach is best suited to surfaces that already run online training. On the infrastructure side there is a direct trade-off between cache capacity and coverage: scaling the hash capacity further raises coverage but requires additional host memory and capacity planning. Finally, the int8 row-wise quantization used for serving efficiency must be handled carefully when int64 feature-cache values are co-located in adjacent structures, which is precisely why we adopt the runtime-metadata design (Appendix~\ref{sec:feature-cache}) rather than a simpler column-appending scheme.

\end{document}